\documentclass[showpacs,aps,prb,twocolumn]{revtex4}

\usepackage{amsbsy,amssymb,amsmath,bm}
\usepackage{graphicx,color,ulem}

\renewcommand{\Re}{\mathop{\mathrm{Re}}}
\newcommand{\f}[1]{Fig.~\ref{#1}} \newcommand{\eq}[1]{Eq.~(\ref{#1})}
\newcommand{\eqs}[2]{Eqs.~(\ref{#1}) and~(\ref{#2})}

 \def\d{\partial} \def\be{\begin{equation}}
\def\ee{\end{equation}} \def\bea{\begin{eqnarray}}
\def\eea{\end{eqnarray}} \def\l({\left(} \def\r){\right)}
\newcommand{\bE}{\mathbf{E}}
 
\newcommand{\Tt}{T_{\text{th}}}
\begin{document}

\title{Dendritic and uniform flux jumps in
superconducting films}

\author{D.~V.~Denisov}
\affiliation{Department of Physics and Center for Advanced
Materials and Nanotechnology,    University of Oslo, P. O. Box
1048 Blindern, 0316 Oslo, Norway}
 \affiliation{A. F. Ioffe Physico-Technical Institute, Polytekhnicheskaya 26, St. Petersburg
194021, Russia}
\author{A.~L.~Rakhmanov}
\affiliation{Institute for Theoretical and Applied
Electrodynamics, Izhorskaya 13/19, Moscow, 125412, Russia}
\author{D.~V.~Shantsev}
\affiliation{Department of Physics and Center for Advanced
Materials and Nanotechnology,    University of Oslo, P. O. Box
1048 Blindern, 0316 Oslo, Norway}
 \affiliation{A. F. Ioffe Physico-Technical Institute, Polytekhnicheskaya 26, St. Petersburg
194021, Russia}
\author{Y.~M.~Galperin}
\affiliation{Department of Physics and Center for Advanced
Materials and Nanotechnology,    University of Oslo, P. O. Box
1048 Blindern, 0316 Oslo, Norway} \affiliation{A. F. Ioffe
Physico-Technical Institute, Polytekhnicheskaya 26, St. Petersburg
194021, Russia}
  \affiliation{Argonne National Laboratory, 9700 S.
Cass Av., Argonne, IL 60439, USA}
\author{T.~H.~Johansen} \email{t.h.johansen@fys.uio.no} \affiliation{Department of Physics and Center
for Advanced Materials and Nanotechnology,   University of Oslo,
P. O. Box 1048 Blindern, 0316 Oslo, Norway}

\date{\today}

\begin{abstract}
Recent theoretical analysis of spatially-nonuniform modes of the
thermomagnetic instability in superconductors \cite{slab} is
generalized to the case of a thin film in a perpendicular applied
field. We solve the thermal diffusion and Maxwell equations taking
into account nonlocal electrodynamics in the film and its thermal
coupling to the substrate. The instability is found to develop in
a nonuniform, fingering pattern if the background electric field,
$E$, is high and the heat transfer coefficient to the substrate,
$h_0$, is small. Otherwise, the instability develops in a uniform
manner. We find the threshold magnetic field,
$H_{\text{fing}}(E,h_0)$, the characteristic finger width, and the
instability build-up time. Thin films are found to be much more
unstable than bulk superconductors, and have a stronger tendency
for formation of fingering (dendritic) pattern.
\end{abstract}


\maketitle

\section{Introduction} \label{In}

The thermomagnetic instability or flux jumping is commonly
observed at low temperatures in type-II superconductors with
strong pinning.~\cite{Mints81, Gurevich97, Wipf91, Wilson83} The
instability arises for two fundamental reasons: (i) motion of
magnetic flux releases energy, and hence increases the local
temperature; (ii)  the temperature rise decreases flux pinning,
and hence facilitates the further flux motion. This positive
feedback can result in thermal runaways and global  flux
redistributions jeopardizing superconducting devices.  The
conventional theory of the thermomagnetic instability
\cite{Mints81, Gurevich97} considers only ``uniform'' flux jumps,
where the flux front is smooth and essentially straight. This
picture is true for many experimental conditions, however, far
from all. Numerous magneto-optical studies have recently revealed
that the thermomagnetic instability in   superconductors can
result in strongly branched dendritic flux
patterns.\cite{Wertheimer67,Leiderer93, Bolz03,
Duran95,welling, Johansen02,Johansen01, Bobyl02,
Barkov03,sust03,ye, Rudnev03,jooss,menghini,Rudnev05}

In a recent paper we examined the problem of flux pattern
formation in the \textit{slab} geometry.~\cite{slab}
Experimentally, however, the dendritic flux patterns are mostly
observed in \textit{thin film} superconductors placed in a
perpendicular magnetic field.  A first analysis of this
perpendicular geometry was recently published by Aranson
\textit{et al.\/}\cite{Aranson} Here we present more exact
and complete picture of the dendritic
instability and analyze the criteria of its realization.

In the following we restrict ourselves to a conventional linear
analysis\cite{Mints81, Gurevich97, Wipf67} of the instability and
consider the space-time development
of small perturbations in the electric field, $E$, and
temperature, $T$. In contrast to the slab case,~\cite{slab} the
heat transfer from the superconductor to a substrate as well as
the nonlocal electrodynamics in thin films are taken into account.
Consequently, the results depend significantly on the heat
transfer rate, $h_0$, as well as on the film thickness, $d$. Our
main result is that the instability in the form of narrow fingers
perpendicular to the background field, $\bE$, occurs much easier
in thin films than in slabs and bulk samples, and the
corresponding threshold field, $E_c$, is found to be proportional
to the film thickness, $d$.

\section{Model and basic equations} \label{Beq}

\begin{figure}
\centerline{\includegraphics[width=8cm]{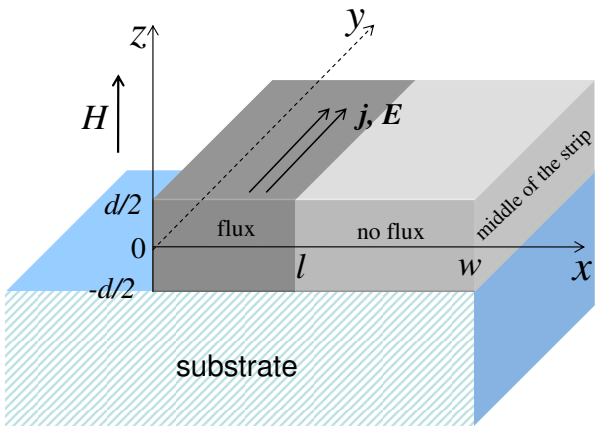}} \caption{ A
superconductor strip on a substrate  (only the left half is
shown). The dark gray area is the flux-penetrated region.
\label{fig:film}}
\end{figure}

Consider the perpendicular geometry shown in Fig.~\ref{fig:film},
with a thin superconducting strip placed in a transverse magnetic
field, $\mathbf{H}$. The strip is infinite along the $y$ axis,
and occupies the space from $-d/2$ to $d/2$ in the $z$-direction
and from $0$ to $2w$ in the $x$-direction. It is assumed that
$d\ll w$. In the unperturbed state the screening current flows
along the $y$-axis .
The distributions of the current density, $\mathbf{j}$, and magnetic
induction, $\mathbf{B}$, in the flux penetrated region $0<x<\ell$ are determined by
the Maxwell equation
\begin{equation}
\label{2} \textrm{curl }\mathbf{B}=\mu_0 \mathbf{j}\, ,
\end{equation}
where the common approximation $\mathbf{B}=\mu_0 \mathbf{H}$ is
used.  To find the electric field and the temperature we use
another Maxwell equation together with the equation for thermal
diffusion,
\begin{eqnarray}
\textrm{curl } \mathbf{E}&=& -\partial \mathbf{B}/\partial t\, , \label{E} \\
C(\partial T/\partial t)&=&\kappa \nabla^2 T+\mathbf{j}\mathbf{E}\, .
\label{th}
\end{eqnarray}
Here $C$ and $\kappa$ are the specific heat and thermal
conductivity, respectively.

Equations (\ref{2})-(\ref{th}) should be supplemented by a
current-voltage relation $j=j(E,B,T)$. For simplicity we assume a
current-voltage curve of the form
\be \mathbf{j} = j_c(T) g(E)\,
(\mathbf{E}/E)\, . \label{cvcf}
\ee
A strong nonlinearity of the function $g(E)$  leads to formation
of a  quasi-static critical state with $j \approx j_c(T)$, where
$j_c$ is the critical current density.\cite{Bean} We neglect any
$B$-dependence of  $j_c$, i. e., adopt the Bean model. The exact
form of $g(E)$ is not crucially important, the only issue is that it represents
a very steep $E(j)$ curve having a large logarithmic derivative,
\be n(E)  \equiv \d \ln E/\d \ln j \approx j_c/\sigma E \gg 1 \, .
\label{n}
\ee
Here $\sigma$ is the differential electrical
conductivity, $\sigma(E)\equiv
\partial j / \partial E$.
The parameter $n$ generalizes the exponent in the frequently used
power-law relation $E \propto j^n$ with $n$ independent of $E$.

The key dimensionless parameter of the model is the ratio of
thermal and magnetic diffusion coefficients,\cite{Mints81}
\begin{equation}\label{tau}
\tau \equiv \mu_0 \kappa \sigma/C \, .
\end{equation}
The smaller $\tau$ is, the slower heat diffuses from the
perturbation region into the surrounding areas. Hence, one can
expect that for smaller  $\tau$: (i) the superconductor is more
unstable, and (ii) the formation of instability-induced nonuniform
structures is more favorable.

In the following we assume that the strip is thinner than the
London penetration depth, $\lambda_L$, and at the same time much
wider than the effective penetration length, $
\lambda_{\text{eff}}=\lambda^2_L/d$,
$$d \leq \lambda_L \ll \sqrt{dw}\, .$$
The stationary current and field distributions in a thin strip
under such conditions were calculated by several
authors,~\cite{Norris70,BrIn,zeld} finding that the flux
penetration depth, $\ell$, is related to the applied field by the expression
\begin{equation}\label{1}
\ell \, /\, w\, =\, \pi^2 H^2/2d^2 j_c^2\ .
\end{equation}
Here it is assumed that the penetration is shallow,
or more precisely that $\lambda_{\text{eff}} \ll \ell \ll w$.


\section{Perturbation analysis} \label{Pan}

\subsection{Linearized dimensionless equations}

We seek solutions of Eqs.~(\ref{2})-(\ref{cvcf}) in the form
\begin{equation} 
T+\delta T(x,y,z,t), \quad \mathbf{E}+\delta \mathbf{E}(x,y,z,t),
\quad \mathbf{j}+\delta \mathbf{j}(x,y,z,t) \nonumber
\end{equation}
where $T$, $\mathbf{E}$ and $\mathbf{j}$ are background values.
The background electric field may be created, e. g., by ramping
the external  magnetic field, and for simplicity we assume it to
be coordinate independent. Allowing for such a dependence would only lead to insignificant
numerical corrections, as discussed
in Ref.~\onlinecite{slab}. Similarly, we will assume a uniform
background temperature.


Whereas it follows from symmetry considerations that $E_x=0$, both
components of the perturbation, $\delta\mathbf{E}$, will in
general not vanish. Linearizing  the current-voltage relation,
Eq.~(\ref{cvcf}) one obtains:
\begin{equation}\label{CVCL}
\delta\mathbf{j}=\left(\frac{\partial j_c}{\partial T}\, \delta T + \sigma \, \delta E_y \right)
\frac{\mathbf{E}}{E}+j_c\frac{\delta\mathbf{E}_x}{E}\, .
\end{equation}
We shall seek perturbations in the form
\begin{eqnarray}
\delta T &=&T^{*}\theta\exp(\lambda t/t_0+ik_x\xi+ik_y\eta)\, , \nonumber 
\\
\delta E_{x,y}&=&E\varepsilon_{x,y}\exp(\lambda t/t_0+ik_x\xi+ik_y\eta)\, , \nonumber 
\\
\delta j_{x,y}&=&j_c i_{x,y}\exp(\lambda t/t_0+ik_x\xi+ik_y\eta)\,
, \label{per3}
\end{eqnarray}
where $\theta$, $\varepsilon$ and $i$ are $z$-dependent
dimensionless Fourier amplitudes.  The
coordinates are normalized to the adiabatic length $a= \sqrt{CT^*/\mu_0 j_c^2}$ where $T^*=-(\partial \ln
j_c/\partial T)^{-1}$ is the characteristic scale of the
temperature dependence of $j_c$, so that $\xi=x/a$, $\eta=y/a$,
$\zeta = z/a$. The time is normalized to $t_0= \sigma C T^*/j_c^2
= \mu_0 \sigma a^2$, which is the magnetic diffusion time for the
length $a$. $\Re \lambda$ is the dimensionless instability
increment, which when positive indicates exponential growth of the
perturbation.
%

We can now use the formulas (\ref{per3}) to rewrite the basic
equations in dimensionless variables. {}From \eq{CVCL} one finds
for the components of the current density perturbation
$\mathbf{i}$,
\begin{equation}\label{i}
 i_x=\varepsilon_x \, , \quad  i_y=-\theta+ n^{-1}\varepsilon_y \, .
\end{equation}
Combining the Maxwell equations ~(\ref{2}) and  (\ref{E}), and the
thermal diffusion equation (\ref{th}) yields
\begin{eqnarray}
&&{\bf k} \times [{\bf k} \times \mathbf{\varepsilon}]=\lambda n\,
\mathbf{i}\, , \label{Eldyn} \\
&&\lambda \theta=\tau
\l(-k_y^2\theta+\frac{\partial^2\theta}{\partial\zeta^2}\r) + (i_y
+ \varepsilon_y)/n \,  . \label{pth}
\end{eqnarray}
Magneto-optical imaging shows that flux patterns produced by the
dendritic instability\cite{Wertheimer67,Leiderer93, Bolz03,
Duran95,welling, Johansen02,Johansen01, Bobyl02,
Barkov03,sust03,ye, Rudnev03,jooss,menghini,Rudnev05} are characterized by having
$k_y \gg k_x$. Therefore, we have neglected the heat flow along
$x$ direction compared to that along the $y$
direction.
Later we will check the
consistency of this assumption by showing that indeed
the fastest growing
perturbation  has $k_y \gg k_x$.

\subsection{Boundary conditions}

We assume that heat exchange between the superconducting film and
its environment follows the Newton cooling law. For simplicity we
let the boundary condition, $\kappa\nabla (T+\delta
T)=-h_0(T+\delta T-T_0)$, apply to both film surfaces. Here
 $T_0$ and $h_0$ are the effective environment temperature
and heat transfer coefficient, respectively. Eqs.~(\ref{pth}) and
~(\ref{i}) can now be integrated over the film thickness to yield
\begin{equation}
\theta=\frac{(1+n^{-1})\varepsilon_y}{n\lambda+n\tau (k_y^2+h)+1}
\nonumber
\end{equation}
where \be h=2h_0 a^2/\kappa d \, . \label{h} \ee In the remaining
part of the paper we let $\theta$, $\mathbf{\varepsilon}$ and
$\mathbf{i}$ denote perturbations
averaged over the film thickness.

 We seek a
solution of the electrodynamic equations in the flux penetrated
region, $0 \le \xi \le \ell/a$.  At the film edge, $\xi=0$, one
has $\delta j_x=0$ and, consequently, $\delta E_x=0$. In the
Meissner state both the electric field and heat dissipation are
absent, so that $\delta E_y=\delta T=\delta j_y=0$ at the flux
front, $\xi=\ell/a$. Thus, the Fourier expansions for the $x$ and
$y$ components of electric field perturbation will contain only
$\sin(k_x\xi)$ and $\cos(k_x\xi)$, respectively. Then the boundary
conditions are satisfied  for
\begin{equation} 
k_x=\left(\pi a/2\ell\right)(2s+1) \, , \quad s=0,1,2,\ldots\,  .
\nonumber
\end{equation}
Since $\ell$ depends on magnetic field, the values of $k_x$ are
also magnetic field dependent.

  Now we can integrate Eq.~(\ref{Eldyn}) over the film
thickness and employ the symmetry of the electrodynamic problem
with respect  to the plane $z=0$. It yields
\begin{eqnarray}
-ik_y(k_x\varepsilon_y+ik_y\varepsilon_x)-\frac{2a}{d}\varepsilon_x'&=&
-\lambda n \varepsilon_x, \nonumber 
\\
-k_x(k_x\varepsilon_y+ik_y\varepsilon_x)+\frac{2a}{d}\varepsilon_y'&=&
-\lambda n f(\lambda, k_y)\varepsilon_y \, .\label{ey1}
\end{eqnarray}
 We have here introduced the function
\begin{equation} 
f(\lambda,k_y) \equiv \frac{i_y}{\varepsilon_y}\frac{1}{n}-\frac{1+n^{-1}}{n\lambda+n\tau (k_y^2+h)+1} \,.
\nonumber
\end{equation}
Note that the  equation for the $z$-component of the field is
satisfied automatically. The derivatives
$\varepsilon_{x,y}^\prime$ with respect to $\zeta$ are taken at
the film surface, $\zeta=d/2a$. To calculate them, one needs the
electric field distribution outside the superconductor, where the
flux density is given by the Bio-Savart law,
\begin{equation}
\mathbf{B(\mathbf{r})}=\mu_0 \mathbf{H} + \frac{\mu_0}{4\pi}\int
d^3\mathbf{r'}\,\frac{\mathbf{j}\times
(\mathbf{r}-\mathbf{r'})}{|\mathbf{r}-\mathbf{r'}|^3}\, .
\nonumber
\end{equation}

The perturbation of flux density is then,
\begin{eqnarray}
&&\delta B_{x,y}=\pm \mu_0 \zeta d \int^{\ell/a}_{0}d\xi'
\int^{\infty}_{-\infty}d\eta'G(\xi-\xi',\eta-\eta') \delta
j_{y,x},\nonumber \\ && \quad
G(\xi,\eta)=\frac{1}{4\pi\left(\xi^2+\eta^2+(d/2a)^2\right)^{3/2}}.
\label{fluxdensity}
\end{eqnarray}
Here we have approximated  the average over $\zeta'$ substituting
$\zeta'=0$. In this way we omit only terms of the order of
$(d/a)^2 \ll 1$.
The integration over $\xi'$ should, in principle, cover also the Meissner region, $\xi'>l/a$.
Though the flux density there remains zero during the development of perturbation, the Meissner current
will be perturbed due to the nonlocal current-field relation.
However the kernel $G(\xi,\eta)$ decays very fast at distances larger than
$d/a$ and therefore the Meissner current perturbation produces
only insignificant numerical corrections.

The perturbation of magnetic field can be related
to that of electrical field by Eq.~(\ref{E}), which can be rewritten as
\begin{equation} \label{eps'}
\delta E'_{x,y}/E=\mp\lambda n\, \delta B_{y,x}/\mu_0 a j_c\,.
\end{equation}
Due to continuity of the magnetic field tangential components
Eq.~(\ref{eps'}) is also valid at the film surface, $\zeta=d/2a$.
Thus it can be substituted  into Eqs.~(\ref{ey1}).
The Fourier components of the kernel function $G(\xi,\eta)$ with
respect to $\eta$ can be calculated directly yielding
\begin{equation}\label{Gk}
G(\xi,k_y)=\frac{k_y a}{2\pi \ell}\cdot\frac{
  \mathrm{K}_1\left(k_y\sqrt{\xi^2+(d/2a)^2}\right)}{
  \sqrt{\xi^2+(d/2a)^2}}
\end{equation}
where $K_1$ is the modified Bessel function of the second kind.

The above Fourier expansions in $\cos(k_x \xi)$ and
$\sin(k_x \xi)$ correspond to the finite interval $-2\ell/a<\xi<2\ell/a$.
Therefore we should continue
$\varepsilon_{x,y}$ from $0<\xi<\ell/a$ to this interval and then introduce $G_x$ and $G_y$ as
analytical continuations of $G(\xi -\xi',k_y)$ having the same
symmetry as  $\varepsilon_x$ and $\varepsilon_y$, respectively (see Section
\ref{app} for details).
All this allows us to rewrite the set (\ref{ey1}) as
\begin{eqnarray}
&& -ik_xk_y\varepsilon_y + (k_y^2+\lambda n)\varepsilon_x
\nonumber \\
&&\qquad =(d/2a)\lambda n\sum_{k_x'} G_x(k_x,k_x',k_y)\varepsilon_x(k_x')\,
,\label{49}\\
&&(k_x^2+\lambda n f)\varepsilon_y+ik_xk_y\varepsilon_x
\nonumber \\
&& \qquad=(d/2a)\lambda n f\sum_{k_x'}
G_y(k_x,k_x',k_y)\varepsilon_y(k_x')\,, \label{50}\\
&& \left\{\begin{array}{c} G_x(k_x,k_x',k_y) \\
    G_y(k_x,k_x',k_y)\end{array}\right\}
=4\int^{\ell/a}_0 \!\! \!\! d\xi \int^{\ell/a}_0\! \!\!\!
d\xi'  G(\xi-\xi',k_y)\nonumber \\
&& \qquad \times \left\{\begin{array}{c}\sin(k_x\xi)\sin(k_x'\xi')  \\
    \cos(k_x\xi)\cos(k_x'\xi')\end{array}\right\} \, . \label{52}
\end{eqnarray}
We are interested only in the
specific case of very thin strip,
\begin{equation}
\label{eq:alpha} \alpha=d/2\ell  \ll 1 \, .
\end{equation}
One can then find analytical expressions for the kernel,
and it turns out that only its diagonal part, $k_x=k_x'$, is
important.


In this manuscript we present analytical expressions up to the
first order in $\alpha$, while the plots are calculated up to the
second order. The second-order analytical expressions can be found
in section \ref{app}. The kernel (\ref{52}) can
be written as
\begin{equation}
 \label{eq:G1de}
G_{x,y}(k_x,k_x,k_y)=\frac{a}{\ell} \left[ \frac{1 - \gamma(\alpha,k_x)\alpha}{\alpha}\right],
\end{equation}
where $\gamma(\alpha,k_x)$ is a dimensionless function.
In what follows we shall consider only the main instability mode, $k_x=\pi a/2 \ell$, which
turns out always to be the most unstable one.
For this mode, and in the limit $\alpha\to 0$, the function $\gamma(\alpha,k_x)$
approaches a constant value $\approx 5$.

 Substituting the above expression for
$G$ into \eqs{49}{50} one obtains the dispersion relation for $\lambda(k_x,k_y)$:
\begin{equation}
    A_1\lambda^2+ A_2\lambda + A_3 =0 \, . \label{eq:de000}
 \end{equation}
Here
\begin{eqnarray}
&& A_1= n \gamma \alpha\, , \quad A_2=k_y^2(1+\tau A_1)+nk_x^2  +A_1
(h\tau-1)\,, \nonumber \\
&&A_3=k_y^4\tau+nk_x^2k_y^2\tau+nk_x^2(h\tau+1/n)+k_y^2(h\tau-1).
\nonumber  
\end{eqnarray}

\section{Results}

\begin{figure}
\centerline{\includegraphics[width=8cm]{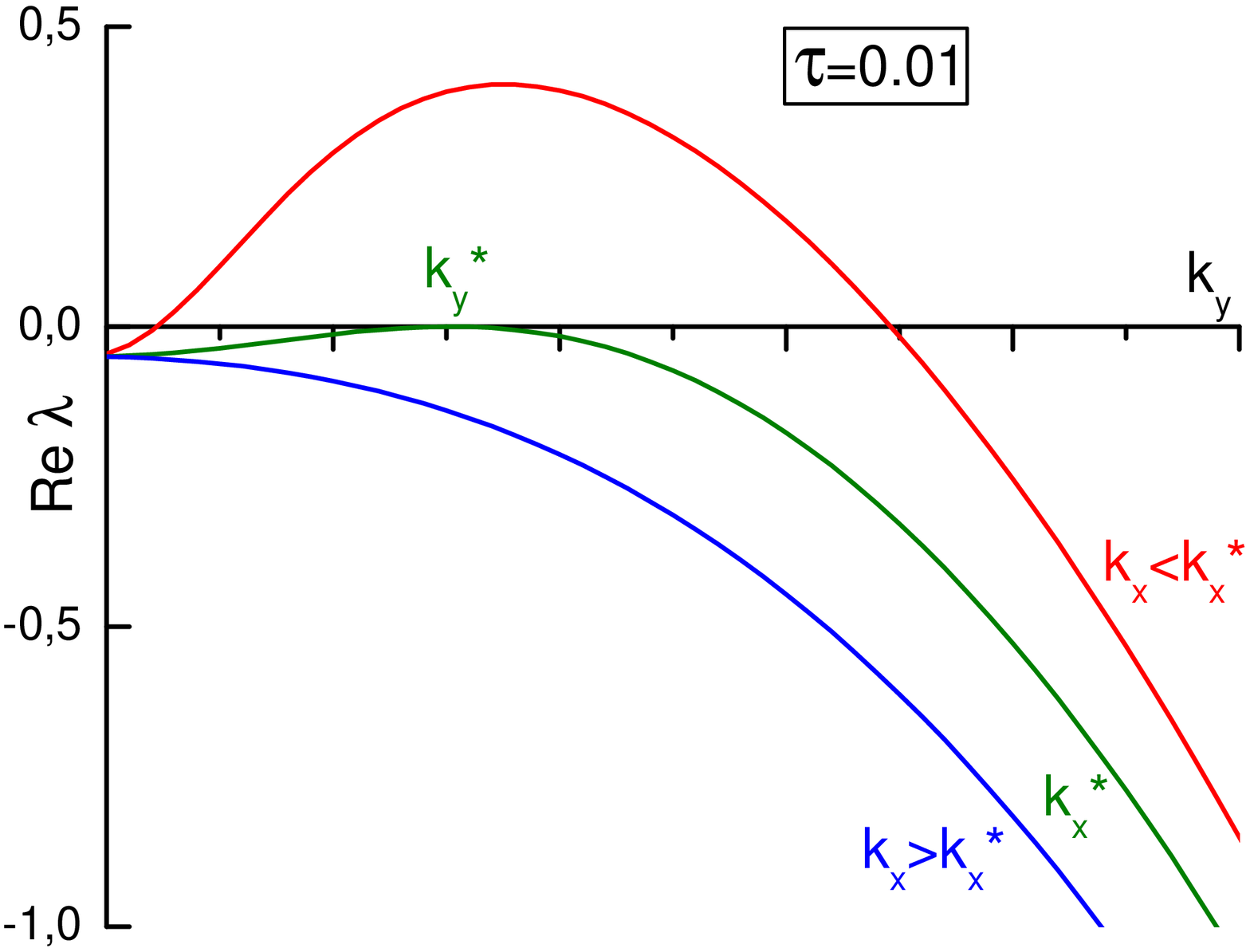}
}\centerline{
\includegraphics[width=8cm]{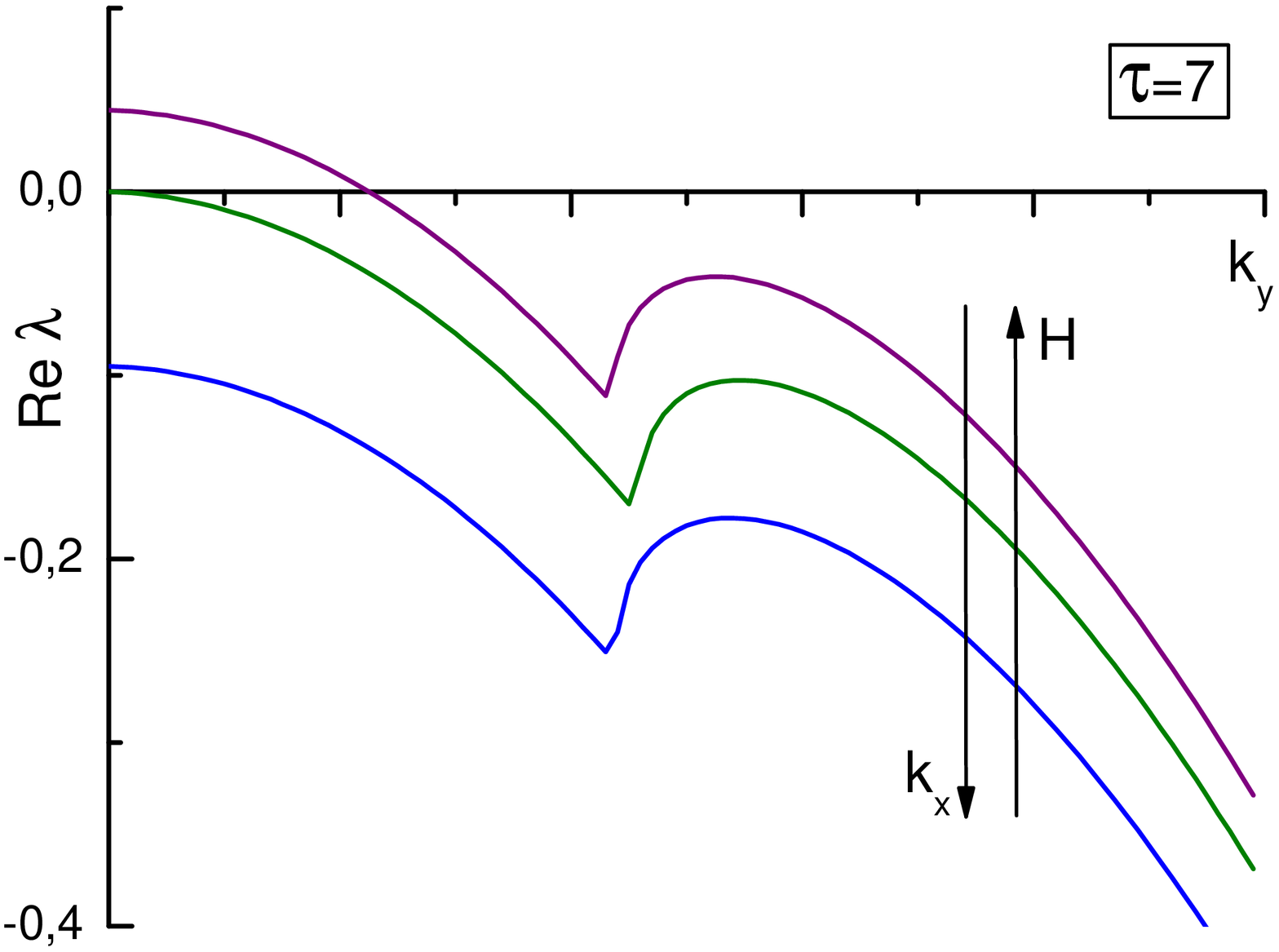}}
\caption{Solutions of dispersion equation \protect{(\ref{eq:de000})} for
small and large  $\tau$, for $\alpha=0.001$ and n=20.  \label{fig:lambda}}
\end{figure}

Let us first consider the simple case of a uniform perturbation, $k_y=0$.
One finds from \eq{eq:de000} that
the perturbation will grow ($\Re \lambda >0$)
 if
\begin{equation}
 \label{ky0}
h\tau < 1 - k_x^2/\gamma \alpha\, .
\end{equation}
When the flux penetration region, $\ell$, is small, i.~e., $k_x$ is large,
the system is stable. As the flux advances,
$k_x$ decreases, and the system can eventually become unstable.
The instability will take place, however, only if $h\tau<1$.
Otherwise the superconducting strip of any width will remain
stable no matter how large magnetic field is applied. This size-independent
stability means that at $h\tau \ge 1$ the heat dissipation due to
flux motion is slower than heat removal into the substrate.

Equation (\ref{ky0}) further simplifies in the adiabatic
limit, $\tau \to 0$, when the heat production is much faster than
heat diffusion within the film or into the substrate. The
instability then develops at $k_x^2/(\gamma \alpha)<1$, which in
dimensional variables  reads as $\mu_0 j_c^2 ld >
CT^*(\pi^2/2\gamma)$. Assuming small penetration depth, $l\ll w$,
and using \eq{1} this criterion can be rewritten as
$H>H_{\text{adiab}}$, with the adiabatic instability field,
\begin{equation}
 \label{H inst}
H_{\text{adiab}} = \sqrt{\frac {d}{w}\ \frac{CT^*}{\gamma\mu_0}}
 \sim \sqrt\frac{d}{w}\, \ H_{\text{adiab}}^{\text{slab}} \, .
\end{equation}
Here $H_{\text{adiab}}^{\text{slab}}$ is the adiabatic
instability field for the slab geometry.\cite{Mints81,Gurevich97,Wipf91,Wilson83,Wipf67}
This result coincides up to a numerical factor with the adiabatic instability field for a thin strip
found recently in Ref. \onlinecite{Shantsev05}.

Solutions of Eq.~(\ref{eq:de000}) for
perturbations with arbitrary $k_y$ are presented in
Fig.~\ref{fig:lambda}. The upper panel shows $\Re \lambda (k_y)$
curves for $\tau = 0.01$ and different values of $k_x$. For large
$k_x$, i.~e., small magnetic field, $\Re \lambda$ is negative  for
all $k_y$. It means that the superconductor is stable. However, at
small $k_x$, the increment $\Re \lambda$ becomes positive in some
finite range of $k_y$. Hence, some perturbations with a spatial
structure will start growing. They will have the form of fingers
of elevated $T$ and $E$ directed perpendicularly to the flux
front. We will call this situation the \textit{fingering (or
dendritic)  instability}.

For large $\tau$ an instability also develops at small $k_x$, however
in a different manner, see Fig.~\ref{fig:lambda} (lower panel).
Here the maximal $\Re \lambda$ always corresponds to $k_y=0$.
Hence, the uniform perturbation will be dominant. The uniform
growth of perturbations for large $\tau$ has been recently
predicted in Refs.~\onlinecite{slab,Aranson} and explained by the
prevailing role of heat diffusion.

Let us now find the critical $k_y^*$ and $k_x^*$ for the fingering
instability, see Fig.~\ref{fig:lambda} (upper
panel). The $k_x^*$ determines the applied magnetic field when the
instability first takes place, while $k_y^*$ determines its
spatial scale. These quantities can be found from the requirement
$\max\{\Re \lambda(k_y)\}=0$ for $k_y\neq 0$. In the limit
$\alpha\ll 1$ we can put  $A_1=0$ in Eq.~(\ref{eq:de000}) and then
rewrite it in the form
\begin{equation}
\lambda=-(k_y^2+h)\tau+\frac{(k_y^2-k_x^2)}{k_y^2+n k_x^2}\, .
\nonumber
\end{equation}
From this expression we obtain
\begin{eqnarray}
k_x^* &=& \left(\sqrt{n+1} - \sqrt{nh\tau}\right)/n\sqrt{\tau} \, ,
\label{k_x*}  \\
k_y^* &=&
\left[\sqrt{nh\tau+1} \left( \sqrt{n+1}-\sqrt{nh\tau+1} \right)\right]^{1/2}/\sqrt{n\tau}
\, . \label{k_y*}
 \nonumber
\end{eqnarray}
%
\begin{figure}
\centerline{
\includegraphics[width=8cm]{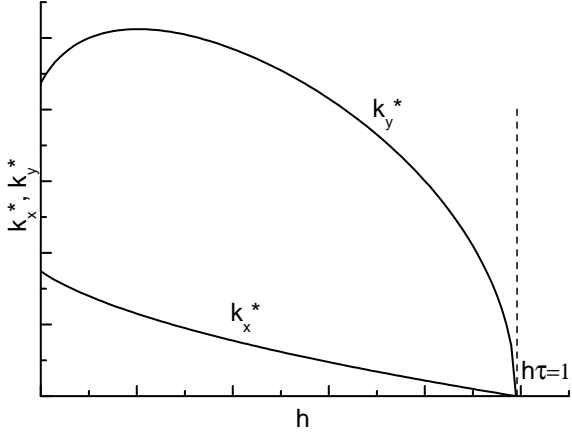}}
\caption{Dependences of $k_y^*$ and $k_x^*$ on $h$ for n=20, $\tau=0.01$, $\alpha=0.001$
according to \eq{k_x*}. \label{fig:k_y&k_x}}
\end{figure}
The dependences of $k_x^*$, $k_y^*$ on the heat transfer coefficient $h$ are shown in
Fig.~\ref{fig:k_y&k_x}.
One can see that $k_y^*$ is always larger than $k_x^*$ implying that
fingers of elevated $T$ and $E$ are extended in the direction normal to the film edge.
For $h\ll 1/\tau$ and $n \gg 1$ we find $k_y^*\approx n^{1/4}k_x^*\gg k_x^* \approx 1/\sqrt{n\tau}$.
Both $k_x^*$ and $k_y^*$ tend to zero as $h \to 1/\tau$, while
for larger $h$ the system is always stable
due to fast heat removal to the substrate.
\begin{figure}
\centerline{
\includegraphics[width=8cm]{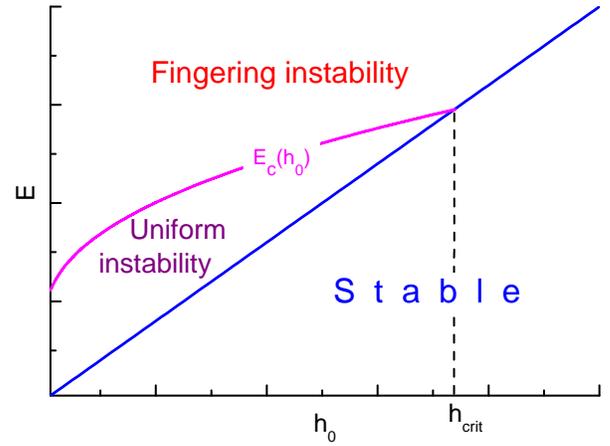}}
\caption{Stability diagram in the plane electric field -- heat transfer
coefficient according to equation \eq{E_c full} and condition $1-h\tau>0$
for $n=30$ and $\alpha=0.001$.  \label{fig:Ec on h0}}
\end{figure}
It follows from \f{fig:lambda} that for large enough $\tau$ the
instability will develop uniformly, while for small $\tau$ it
will acquire a spatially-nonuniform structure. Let us find now the
critical value $\tau_c$ that separates these two regimes. It can
be obtained from the equality $\Re \lambda(k_x=k_x^*,k_y=0)=0$.
When it is fulfilled $\Re \lambda=0$ both for $k_y=0$ and for
$k_y=k_y^*\ne 0$. We find using Eq.~(\ref{eq:de000}) that the
instability will evolve in a spatially-nonuniform way if
\begin{eqnarray}
 \label{tau crit}
\tau<\tau_c=\left(1-k_x^{*2}/\gamma \alpha\right)/h \, .
\end{eqnarray}
Substituting here $\alpha$ and  $k_x^*$ we find a transcendental relation
between $\tau_c$ and $h$. 
For $n\gg 1$ it reduces to
\begin{equation}
\label{E_c full} \sqrt{n \tau_c}\left(1+\sqrt{h\tau_c} \right)=\pi
a/\gamma d
 \, .
\end{equation}
Using this result we can
construct a stability diagram in the $E-h_0$ plane shown in
Fig. \ref{fig:Ec on h0}.
The curved line marks the critical electric field $E_c(h_0)$
that separates
two types of instability: fingering ($E>E_c$) and uniform ($E<E_c$).
This line is calculated from \eq{E_c full}, where the electric field is expressed via
$\tau$ as $E=j_c\mu_0\kappa /nC \tau$ according to \eqs{n}{tau}.
The straight line is given by the condition $h\tau=1$.
Below this line the superconductor is always stable,
as follows from
\eq{ky0} for the uniform perturbations, and from \eq{k_x*} for the nonuniform case.
At a certain value $h_0 = h_{\text{crit}}$, the two lines intercept.
We find
\begin{equation}
\label{h_crit} h_{\text{crit}}=\frac{2 \gamma^2 \mu_0^2 j_c^4 d^3
\kappa n}{\pi^2 T^{* 2} C^2}\, ,
\end{equation}
and the critical electric field $E_c$ for $h_0=0$ is
\begin{equation}
\label{critelecfield}
E_{c}(0)=\frac{\gamma^2 \mu_0^2 \kappa j_c^3}{\pi^2 C^2 T^*}\ d^2 \ ,
\end{equation}
while $E_c(h_{\text{crit}})=4 E_c(0)$.
\begin{figure}[t]
\centerline{\includegraphics[width=8cm]{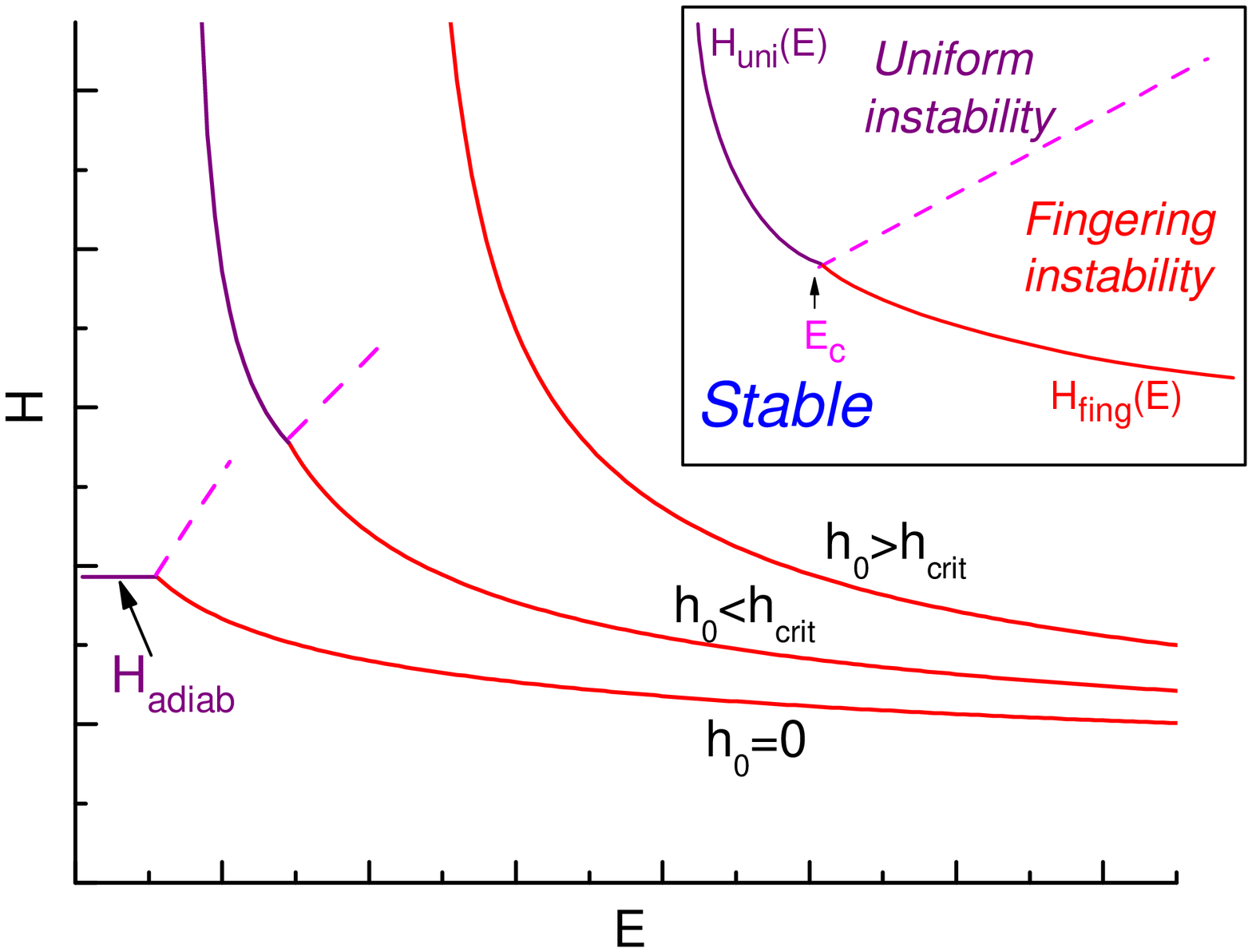}}
\caption{Stability diagram in the $H$-$E$ plane according to \eq{ky0} and \eq{k_x*}. \label{fig:phase}}
\end{figure}

For any point ($h_0$,$E$) belonging to the stable phase in the
stability diagram, \f{fig:Ec on h0}, the flux distribution is
stable for any applied magnetic field. For the points belonging to
unstable phases, the instability develops above some threshold
magnetic field, either $H_{\text{fing}}(h_0,E)$ or
$H_{\text{uni}}(h_0,E)$ for fingering or uniform instability,
respectively. Shown in Fig.~\ref{fig:phase} are three sets of
$H_{\text{fing}}(E)$ and $H_{\text{uni}}(E)$ curves for different
values of $h_0$. They represent boundaries between the three
phases, stable and unstable with respect to either fingering or
uniform instability, as shown in the inset. Using \eq{1} one can
rewrite the expression \eq{ky0} for $H_{\text{uni}}$ as
\be
 H_{\text{uni}} = H_{\text{adiab}}
 \left( 1- \frac{2 T^* h_0 }{nd\, j_c E}\right)^{-1/2} \, .
 \label{uni}
\ee
In the absence of heat removal to the substrate, $h_0=0$, we obtain the adiabatic instability field, \eq{H
inst}, and the $H_{\text{uni}}(E)$ curve becomes a horizontal
line.\cite{heatx}

The threshold magnetic
field for the \textit{fingering} instability, $H_{\text{fing}}$, is calculated from \eq{k_x*}.
A simplified expression obtained for $h\ll 1/\tau$ and $n\gg 1$,
\begin{equation}
H_{\text{fing}}=\left(\frac{j_c d^2}{\pi w} \sqrt{\frac{\kappa T^*
j_c} {E}} \right)^{1/2}\, , \label{Hfing}
\end{equation}
shows that at large electric fields
$H_{\text{fing}}$ decays as $E^{-1/4}$.
At $h_0 \le h_{\text{crit}}$ the curves $H_{\text{fing}}(E)$ and
$H_{\text{uni}}(E)$ intercept at the critical electric field $E_c$
determined by \eq{tau crit}.
At $h_0 \ge h_{\text{crit}}$ we have
$H_{\text{fing}}(E)<H_{\text{uni}}(E)$ for any $E$, so the lines do
not intercept and the instability will develop into a fingering
pattern.

\section{Discussion}

Let us compare the present results for a \textit{thin film} in a
perpendicular magnetic field with results of
Ref.~\onlinecite{slab} for a \textit{bulk} superconductor. In both
cases the instability develops into a fingering pattern if the
background electric field in the superconductor exceeds some
critical value $E_c$. The values of $E_c$ are however different.
Their ratio for a thin strip and a slab, \be
\frac{E_c(0)}{E_c^{\text{slab}}} = \frac{\gamma^2}{\pi^2}\
\frac{d^2 j_c^2 \mu_0}{CT^*}\, , \ee is expected to be much less
than unity. For $j_c=10^{10}$ A/m$^2$, $C=10^3$ J/Km$^3$,
$\kappa=10^{-2}$ W/Km, $T^*=10$ K and $d=0.3\, \mu$m, we find from
\eq{critelecfield} that $E_c\approx 4\cdot 10^{-4}$~V/m, while
according to Ref.~\onlinecite{slab}, $E_c^{\text{slab}}=0.1$ V/m.
Consequently, the development of thermomagnetic instability into a
fingering pattern is much more probable in thin films than in bulk
superconductors.

The threshold \textit{magnetic} field for the fingering
instability, $H_{\text{fing}}$, is also much smaller for thin
films. Comparing \eq{Hfing} with the results of
Ref.~\onlinecite{slab} for a slab\cite{slab} we find
\begin{eqnarray}
\frac{H_{\text{fing}}}{H_{\text{inst}}^{\text{slab}}} = 
\frac{\sqrt{2}}{\pi} \frac{d}{\sqrt{w l^*}} \, . \label{ch}
  \end{eqnarray}
Here $l^*=(\pi/2)\sqrt{\kappa T^*/j_cE}$ is the flux penetration
depth at the threshold of the fingering instability,
$H=H_{\text{fing}}$. Experimentally, the fingering instability
always starts after the flux has penetrated a noticeable distance
from the edges, such that $l^*\gg
d$.\cite{Wertheimer67,Leiderer93, Bolz03, Duran95,welling,
Johansen02,Johansen01, Bobyl02, Barkov03,sust03,ye,
Rudnev03,jooss,menghini,Rudnev05}
Hence, for a thin film the fingering instability
should start at much smaller applied fields than in bulk
samples (by a factor of $\sim 10^3$ for films with $d\sim 10^{-4}w$). The difference between the
threshold fields for the two geometries here is even stronger than
for the case of uniform instability in the adiabatic limit, see
\eq{H inst}. Assuming the above values of parameters and $w=2$~mm
we find from \eq{Hfing} that
$H_{\text{fing}}[E_c(0)]=H_{\text{adiab}}\approx 1$~mT. This value
becomes larger if we take into account the heat transfer to the
substrate. It is therefore in excellent agreement with
experiment,\cite{Duran95,welling,Johansen02,
Barkov03,Rudnev03,jooss,Rudnev05,Shantsev05} where the threshold
field is typically of the order of a few milliTesla.

The spatial structure of the instability predicted by our linear analysis
is a periodic array of fingers perpendicular to the film edge.
Its period  can be estimated from \eq{k_x*}.
For $E=E_c$, $h=0$  and $n\gg 1$ one finds
\begin{equation}
\label{ell}
d_y=\frac{\pi^2 CT^*}{2\gamma  n^{1/4} \mu_0 j_c^2  d}\, ,
\end{equation}
which yields $d_y \approx 100\ \mu$m for $n=30$. Numerical
analysis of the instability development shows\cite{Aranson,slab}
that beyond the linear regime the periodic structure is destroyed
and only one (strongest) finger invades the Meissner region. This
scenario is indeed reproduced experimentally, and the observed
width of individual fingers, 20-50~$\mu$m,\cite{Leiderer93,
Duran95, Johansen01, Barkov03} is very close to our estimate,
$d_y/2$.

The finger width and the threshold magnetic field also depend on
the dimensionless parameter $h$ characterizing the thermal
coupling to the substrate, \eq{h}. In turn, $h$, grows rapidly
with temperature because of a strong $T$ dependence of $C$ and
$j_c$. One can therefore make several testable predictions from
the dependences $k_x^*(h)$ and $k_y^*(h)$ shown in
Fig.~\ref{fig:k_y&k_x}: (i) There must be a threshold temperature
$\Tt$ above which the instability is not observed. (ii) When
approaching $\Tt$, the instability field diverges since $k_x^* \to
0$. (iii) When approaching $\Tt$ the characteristic width of
individual fingers increases since $k_y^* \to 0$. 
The last prediction has also been obtained in the boundary layer model
allowing calculation of the exact finger shape.\cite{Baggio} 
The first and
the second predictions have already been confirmed
experimentally.\cite{Johansen01,welling} 
As for the last one, the
$T$ dependence of the finger width has not yet been studied. At
the same time, there is a solid experimental
evidence\cite{Duran95, Johansen01, welling, Rudnev05} for an
enhanced degree of branching as $T \to \Tt$ that can be
quantitatively described as a larger fractal dimension of the flux
pattern.\cite{Rudnev05} This abundant branching could be an
indirect consequence of the increased finger width since wider
fingers are presumably more likely to undergo splitting.

The present problem of fingering instability in a thin film has
two new features compared to  a similar problem for a bulk
superconductor, (i) nonlocal electrodynamics and (ii) thermal
coupling to the substrate. The nonlocality results in much smaller
values of the threshold magnetic field $H_{\text{fing}}$ and the
critical electric field $E_c$
in films than in bulks. 
If a film is made thinner, it becomes even more unstable since $H_{\text{fing}}\propto d$, and
has a stronger tendency to form a fingering pattern since $E_c(0)\propto d^2$.
The thermal coupling to the substrate has a somewhat opposite effect.
It can lead to an ultimate stability
if $h>1/\tau$ -- a situation that is never realized in bulks.
A moderate coupling, $h\ll 1/\tau$, slightly renormalizes $H_{\text{fing}}$ and $E_c$, i.e.
makes the film a little bit more stable and less inclined to fingering.

Let us now compare the results presented in this work to those
obtained in a similar model by  Aranson 
\textit{et al}.\cite{Aranson}
Our expressions for the ``fingering'' threshold field 
, \eq{Hfing}, and for the finger width, \eq{ell}, agree with their
results up to a numerical factor.
For $\tau \gg 1$ our results for the ``uniform''  threshold field
(derived from \eq{ky0}) are also similar to results of Ref.~\onlinecite{Aranson}.
As a new result,  we
find that there exists a critical value of the parameter $\tau$,
\eq{tau crit}, which controls whether the instability evolves
either in the uniform, or in the fingering way. Shown in \f{fig:Ec
on h0} is the stability diagram where the line $E_c(h_0)$
separates regimes of fingering and uniform instability. Other new
results of this paper are: (i) the existence of a
field-independent ``critical point", $h_{\text{crit}}$, such that
for $h_0> h_{\text{crit}}$ the instability \textit{always}
develops into a fingering pattern, and (ii) the full  stability
diagram in the $H$-$E$ plane, \f{fig:phase}, containing all three
phases.

The background electric field  needed to nucleate the fingering
instability can be  induced by ramping the magnetic field, $E\sim
\dot{H} l \propto \dot{H} H^2$ for $l\ll w$, where $\dot{H}$ is
the ramp rate. This is the lowest estimate since the flux
penetration in practice is strongly nonuniform in space and in
time,\cite{AltRev} and there can be additional sources of $E$ due
to random fluctuations of superconducting parameters. The
occurrence of the fingering instability even at rather low ramp rates
\cite{Wertheimer67, Duran95,welling, Johansen02, Barkov03,
Rudnev03,jooss,menghini,Rudnev05} is therefore not surprising.

The build-up time of the instability can be estimated as $t_0 \approx 0.1\mu$s~
if the flux-flow conductivity $\sigma = 10^9\ \Omega^{-1}m^{-1}$.
Our linear analysis assumes that the perturbations of $T$ and $E$  grow in amplitude, but remain localized
within the initial flux penetrated region.
Numerical results show\cite{slab,Aranson} that at $t\gg t_0$
the perturbations also propagate into the Meissner region.
This propagation can be described by recent models\cite{biehler05,shapiro}
that predict a characteristic propagation speed in agreement
with experimental values of 10-100~km/s.\cite{Leiderer93,Bolz03}

In conclusion, the linear analysis of thermal diffusion and
Maxwell equations shows that a thermomagnetic instability in a
superconducting film may result in either uniform or finger-like
distributions of $T$, $E$ and $B$. The fingering distributions
will be  observed if the background electric field $E>E_c$, where
$E_c$ grows with the film thickness, the critical current density,
the thermal conductivity and the thermal coupling to the
substrate. Due to nonlocal electrodynamics in thin films they turn
out to be more unstable than bulk superconductors and more
susceptible to formation of a fingering pattern.

\begin{acknowledgments}
This work is supported by the Norwegian Research Council, Grant.
No. 158518/431 (NANOMAT), by the U.~S. Department of Energy
Office of Science through contract No. W-31-109-ENG-38 and
by Russian Foundation for Basic
Research through project No. 03-02-16626. We are
thankful for helpful discussions with V. Vinokur, A. Gurevich, and
I. Aranson, U. Welp, R. Wijngaarden, and V. Vlasko-Vlasov.

\end{acknowledgments}

\widetext
\newpage
\section{Appendix} \label{app}

Here we present in  more detail some derivations  omitted in main text.

\subsection{\eq{Eldyn}}
In components, equation (\ref{Eldyn}) reads as
\begin{eqnarray}
ik_y\left(\frac{\partial
\varepsilon_y}{\partial\xi}-ik_y\varepsilon_x\right)-\frac{\partial^2\varepsilon_x}{\partial\zeta^2}&=&-\lambda
n \varepsilon_x\, , \label{ex}\\
\frac{\partial}{\partial\xi}\left(\frac{\partial
\varepsilon_y}{\partial\xi}-ik_y\varepsilon_x\right)+\frac{\partial^2\varepsilon_y}{\partial\zeta^2}&=&\lambda
n f(\lambda, k_y)\varepsilon_y\, ,\label{ey} \\
\frac{\partial}{\partial\zeta}\left(\frac{\partial\varepsilon_x}{\partial\xi}+ik_y\varepsilon_y\right)&=&0
\, .\label{ez}
\end{eqnarray}

\subsection{Transition from Eq.(\ref{Gk}) to Eq.(\ref{52})}

The Fourier transformation for the kernel function $G$ with respect to variable $\eta$ reads
\begin{eqnarray}
G(\xi,\eta,\alpha)&=&\int^{\infty}_{-\infty}\frac{dk_y}{2\pi}G(\xi,k_y,\alpha)e^{ik_y\eta}\,,\\
G(\xi,k_y,\alpha)&=&\int^{\infty}_{-\infty}d\eta
G(\xi,\eta,\alpha)e^{-ik_y\eta} \, ,
\end{eqnarray}

Substituting Fourier transformation of kernel $G$ into Eqs. (\ref{fluxdensity}) and (\ref{eps'}), we find
\begin{eqnarray}
\varepsilon'_x(\xi,k_y)=2\alpha^2\lambda n\int^{\ell/a}_{0}d\xi' G(\xi-\xi',k_y)\varepsilon_x(\xi',k_y),\quad \label{34}\\
\varepsilon'_y(\xi,k_y)=2\alpha^2\lambda n f(\lambda,k_y)\int^{\ell/a}_{0}d\xi' G(\xi-\xi',k_y)\varepsilon_y(\xi',k_y).\quad\label{35}
\end{eqnarray}
These are two linear independent integral equations with difference kernel. 
Each of these equations can be solved separately by a standard method.

Although the procedure of solution of Eqs. (\ref{34}) and (\ref{35}) is standard, 
we describe it in details to avoid any mistakes. In a strict sense, 
the symmetry of the Fourier expansions with $\cos(k_x \xi)$ and $\sin(k_x \xi)$ 
corresponds to the interval $-2\ell/a<\xi<2\ell/a$ for these trigonometric functions 
to be orthogonal at different $k_x$. Thus, we should formally continue 
analytically $\varepsilon_{x,y}$ on this interval. These continuations are 
different for the sine and cosine functions. It could be easily found that the symmetry of the function $\varepsilon_{y}(\xi)$ corresponds to the 
following continuation $\widetilde{\varepsilon}_{y}(\xi)$ from the interval $0<\xi<\ell/a$ to $-2\ell/a<\xi<2\ell/a$
\begin{equation}\label{36}
\widetilde{\varepsilon}_{y}(\xi)=\varepsilon_{y}(2\ell/a-\xi)\,\,
\text{at} \,\, \ell/a<\xi<2\ell/a
\end{equation}
and $\widetilde{\varepsilon}_{y}(\xi)$ is an odd function with respect to $\xi=0$. The continuation $\widetilde{\varepsilon}_{x}(\xi)$ of the 
function $\varepsilon_{x}(\xi)$ is
\begin{equation}\label{37}
\widetilde{\varepsilon}_{x}(\xi)=-\varepsilon_{x}(2\ell/a-\xi)\,\,
\text{at} \,\, \ell<\xi<2\ell/a
\end{equation}
and is even with respect to $\xi=0$. The functions $\widetilde{\varepsilon}_{x,y}(\xi)$ 
coincide with $\varepsilon_{x,y}(\xi)$ at $0<\xi<\ell/a$ and can be expanded 
using the same trigonometrical functions. So, we get
\begin{eqnarray}
\varepsilon_x(k_x)=\frac{1}{2}\int^{2\ell/a}_{-2\ell/a}\widetilde{\varepsilon}_x(\xi)\cos(k_x\xi)d\xi=2\int^{\ell/a}_0\varepsilon_x(\xi)\cos(k_x\xi)
d\xi,\,\,\, \label{38}\\
\varepsilon_y(k_x)=\frac{1}{2}\int^{2\ell/a}_{-2\ell/a}\widetilde{\varepsilon}_y(\xi)\sin(k_x\xi)d\xi=2\int^{\ell/a}_0\varepsilon_y(\xi)\sin(k_x\xi)
d\xi.\,\,\, \label{39}
\end{eqnarray}

Let us produce a similar continuations with the function $G$.
First, we consider $G(\xi-\xi')$ in Eq. (\ref{34}) as a function of two
independent arguments, $G(\xi,\xi')$. 
Then we can define the continuation $\widetilde{G}_x(\xi')$ of $G(\xi')$ as 
\begin{equation}\label{40} 
\widetilde{G}_x(\xi')=-G(2\ell/a-\xi')\,\, \text{at} \,\, \ell/a<\xi'<2\ell/a
\end{equation}
and even with respect to $\xi'=0$. Second, in Eq. (\ref{35}), we can define the continuation $\widetilde{G}_y(\xi')$ of
$G(\xi')$ as
\begin{equation}\label{41}
\widetilde{G}_y(\xi')=G(2\ell/a-\xi')\,\, \text{at} \,\, \ell/a<\xi'<2\ell/a
\end{equation}
and $\widetilde{G}_y(\xi')$ is an odd function with respect to
$\xi'=0$. The functions $\widetilde{G}_{x,y}(\xi,\xi')$ obey the same
symmetry with respect to the other argument $\xi$ since
$G[(\xi-\xi')]$ is symmetric with respect to a permutation of
$\xi$ and $\xi'$ as follows from (\ref{fluxdensity}).

The integrals, Eqs. (\ref{34}) and (\ref{35}), can now be written as
\begin{eqnarray}
\widetilde{\varepsilon}_x\,'(\xi)=\frac{\alpha^2\lambda n}{2}\int^{2\ell/a}_{-2\ell/a}d\xi' \widetilde{G}_x(\xi,\xi')\widetilde{\varepsilon}_x(\xi') 
,\,\,\,\label{42}\\
\widetilde{\varepsilon}_y\,'(\xi)=\frac{\alpha^2\lambda n f(\lambda,k_y)}{2}\int^{2\ell/a}_{-2\ell/a}d\xi' 
\widetilde{G}_y(\xi,\xi')\widetilde{\varepsilon}_y(\xi'). \,\,\,\label{43}
\end{eqnarray}
The solutions of these equations coincide with the solutions of
the original Eqs. (\ref{34}) and (\ref{35}) at $0<\xi<\ell/a$ and have the
symmetry necessary for the suggested Fourier transformations.

Let us introduce the Fourier transformations
\begin{eqnarray}
\widetilde{G}_x(\xi,\xi')=\sum_{k_x,k_x'}G_x(k_x,k_x')\cos(k_x\xi)\cos(k_x'\xi'),\label{44}\\
\widetilde{G}_y(\xi,\xi')=\sum_{k_x,k_x'}G_y(k_x,k_x')\sin(k_x\xi)\sin(k_x'\xi'),\label{45}
\end{eqnarray}
where
\begin{eqnarray}
G_x(k_x,k_x')=4\int^{\ell/a}_0d\xi\int^{\ell/a}_0 d\xi'G_x(\xi,\xi')\cos(k_x\xi)\cos(k_x'\xi'),\quad \label{46}\\
G_y(k_x,k_x')=4\int^{\ell/a}_0d\xi\int^{\ell/a}_0 d\xi'G_y(\xi,\xi')\sin(k_x\xi)\sin(k_x'\xi'),\quad \label{47}
\end{eqnarray}
and $G_{x,y}(k_x,k_x')=G_{x,y}(k_x',k_x)$. Equations (\ref{46}),(\ref{47}) lead to (\ref{52})

\subsection{Properties of kernel $G$}

First, note that we do not disregard the value of the order of $(d/\ell)^2$ in 
the denominator of the function $G$. It seems as an exceeding of accuracy 
since previously we neglect the $z$ dependence of electric field and temperature across the film. 
In other words, in the film
\begin{equation}\label{51}
\varepsilon_i(\zeta)=\varepsilon_i(0)+\frac{\zeta^2}{2}\frac{\partial^2\varepsilon_i}{\partial\zeta^2}+...\,,
\end{equation}
where the first derivative is omitted due to the symmetry. We neglect all the terms in the right hand side of Eq. (\ref{51}), except the first. That 
is, we should seemingly neglect all the values of the order of $(d/\ell)^2$ and for the temperature perturbation as well. However, it is not so 
since the derivatives across the film thickness includes additional smallness. In the case of electromagnetic values it is due to the assumption 
that $d\leq\lambda_L$ and for the temperature due to the smallness of Bio number $Bi=h_0d/\kappa$ in any realistic situation for a thin
film.

As in the previous study \cite{slab}, we analyze here the stability in the 
linear approximation following two different ways. In qualitative approach, we do 
not specify the boundary conditions exactly assuming that $k_x$ is of the order of $a/\ell$, 
and analyze evolution of perturbations with each $k_x$ independently. In this approach we evidently need only diagonal component of the kernel $G$ 
with $k_x=k_x'$. If we specify the boundary conditions and try to find exact expressions for the stability criterion and characteristic time and 
spatial scales, we need to know kernel function in a more general form. Both the approaches should give rise to qualitatively the same results. 

\subsection{Transition from \eqs{49}{52} to formula (\ref{eq:G1de})}

Performing the integration of Fourier transformation for the kernel function $G$ we get
\begin{equation}\label{Gk2}
G(\xi,k_y)=\frac{k_y a}{2\pi \ell \sqrt{\xi^2+\alpha^2}}
K_1\left(k_y\sqrt{\xi^2+\alpha^2}\right),
\end{equation}
where $K_1$ is the modified Bessel function of the second kind. In general case, the kernel function $G(k_x,k_x',k_y)$ could not be found in the 
explicit form and Eqs. (\ref{49}) and (\ref{50}) should be solved numerically. However, we are interested here only in the specific case $\alpha \ll 
1$. Within these limit the analytical expressions for the kernel can be found.

The expressions for the Fourier components of the function $G$ read
\begin{eqnarray}
G_x(k_x,k_x',k_y)=\frac{2k_y a}{\pi \ell}\int^{\ell/a}_0 d\xi \int^{\ell/a}_0 d\xi'\frac{K_1\left[k_y\sqrt{(\xi-\xi')^2+\alpha^2}\right]
\cos(k_x\xi)\cos(k_x'\xi')}{\sqrt{(\xi-\xi')^2+\alpha^2}},\label{522} \\
G_y(k_x,k_x',k_y)=\frac{2k_y a}{\pi \ell}\int^{\ell/a}_0 d\xi \int^{\ell/a}_0 d\xi'\frac{K_1\left[k_y\sqrt{(\xi-\xi')^2+\alpha^2}\right]
\sin(k_x\xi)\sin(k_x'\xi')}{\sqrt{(\xi-\xi')^2+\alpha^2}}.\label{53}
\end{eqnarray}

Under conditions specified for (\ref{522}),(\ref{53}), the main contribution to these integrals evidently comes from the region $|\xi-\xi'|<\alpha$. 
At larger $|\xi-\xi'|$ the functions under integrals decays exponentially. So, we can replace the Bessel function $K_1(x)$ under integrals by its 
expression at small value of argument $x$ up to the order of $x$ 
\begin{equation}
\label{ass}
K_1(x)\approx \frac{1}{x}+\frac{x}{2}\ln x
\end{equation}

From Eqs. (\ref{522}) and (\ref{53}) we find in the main
approximation accounting for the first term in the rhs of Eq.
(\ref{ass})
\begin{eqnarray}
G^1_x(k_x,k_x',k_y)=\frac{2 a}{\pi \ell}\int^{\ell/a}_0 d\xi\int^{\ell/a-\xi}_{-\xi} du\frac{\cos(k_x\xi)\cos[k_x'(\xi+u)]}{u^2+\alpha^2},\label{54} 
\\
G^1_y(k_x,k_x',k_y)=\frac{2 a}{\pi \ell}\int^{\ell/a}_0 d\xi\int^{\ell/a-\xi}_{-\xi} du\frac{\sin(k_x\xi)\sin[k_x'(\xi+u)]}{u^2+\alpha^2},\label{55}
\end{eqnarray}
where we proceed to integration over $u=\xi'-\xi$.

We need to compute following integrals, the first one for $G^1_x$ and the second for $G^1_y$:
\begin{eqnarray}
\int^{\ell/a}_0 \cos k_x\xi \cos k_x'\xi d\xi \int^{\ell/a-\xi}_{\xi} du \frac{\cos k_x'u}{u^2+\alpha^2} -
\int^{\ell/a}_0 \cos k_x\xi \sin k_x'\xi d\xi \int^{\ell/a-\xi}_{\xi} du \frac{\sin k_x'u}{u^2+\alpha^2}
\end{eqnarray}
\begin{eqnarray}
\int^{\ell/a}_0 \sin k_x\xi \sin k_x'\xi d\xi \int^{\ell/a-\xi}_{\xi} du \frac{\cos k_x'u}{u^2+\alpha^2} +
\int^{\ell/a}_0 \sin k_x\xi \cos k_x'\xi d\xi \int^{\ell/a-\xi}_{\xi} du \frac{\sin k_x'u}{u^2+\alpha^2}
\end{eqnarray}
The calculation of this integrals are almost identical, so we show only computing of the first one. 
In the begining we need to compute following integral

\begin{eqnarray}
\int_{-\xi}^{\ell/a-\xi}\! \!  d u\,  \frac{\cos k_xu}{u^2+\alpha^2} =
\int_{-\xi}^{\ell/a-\xi} \! \! \frac{ d u}{u^2+\alpha^2} 
 - \int_{-\xi}^{\ell/a-\xi} d u\, \frac{1 - \cos k_xu }{u^2+\alpha^2} 
 \, . \label{eq:tmp1}
\end{eqnarray}
The first integral yields $\alpha^{-1}\left[\arctan (\xi/\alpha) + \arctan
  (\ell/a-\xi)/\alpha \right]$. Thus we have to calculate
$$\frac{2 a}{ \pi \ell}\int_0^{\ell/a} d\xi\, \cos k_x\xi \cos k_x'\xi \left[\arctan (\xi/\alpha) + \arctan
  (\ell/a-\xi)/\alpha \right] \, .$$
One can show that off-diagonal integrals, i. e. for $k_x \ne k_x'$, are
  very small, and we will keep only diagonal elements which are the
  same for all $k_x$. Let us denote them as
  $[1-c(\alpha)]/\alpha$. For $\alpha\ll 1$ one finds $c(\alpha) \propto \alpha$,
while  a good approximation for $\alpha \lesssim 1$ case is  
  $c(\alpha) \approx 1.6a^{0.84}$.
Even though $c(\alpha) \to 0$ for $\alpha \to 0$ 
its account is important since $c(\alpha)$
  will enter the dispersion law with the factor $\propto k^2$. 

The second integral in Eq.~(\ref{eq:tmp1}) can be calculated numerically, it
has the order of $k_x^2$, let us denote this integral as $r(\alpha,k_x)$.
The values of $r(\alpha,k_x)$ in the limit $\alpha\ll 1$ depend on $k_x$,
namely $r(\alpha,k_x)=0$ for $k_x=0$ and $r(\alpha,k_x) \to k_x$ for $k_x \gg 1$. 
 
To complete computation of Eqs. (\ref{54}) and (\ref{55}) we also need to calculate integral
\begin{eqnarray}
\int^{\ell/a}_0 \cos k_x\xi \sin k_x'\xi d\xi \int^{\ell/a-\xi}_{\xi} du \frac{\sin k_x'u}{u^2+\alpha^2},
\end{eqnarray}
which for $k_x=(\pi a/2\ell)(2n+1)$ is equal to zero. The same results one can obtain for $G_y^0$. Despite the fact that Eq. (\ref{54}) contains 
$cos$ and Eq. (\ref{55}) contains $sin$ the final results are the same thus we employ the approximation

\begin{equation}
  \label{eq:app01}
G_{x,y}^1(k_x,k_x;k_y)=\frac{a}{\ell}\left[\frac{1-c(\alpha)}{\alpha} - r(\alpha,k_x)\right]
\end{equation}  

\begin{figure}
\centerline{\includegraphics[width=8cm]{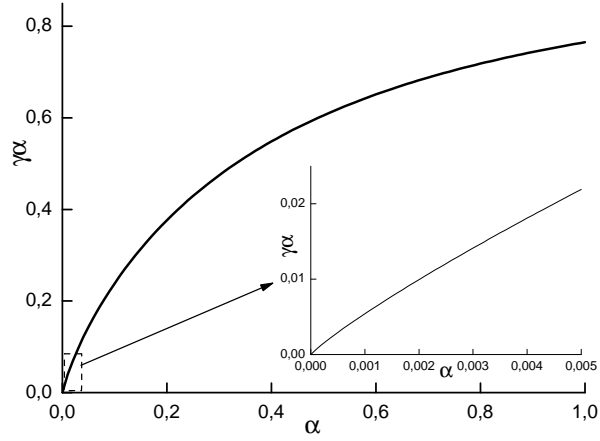}}
\caption{Plot of the product $\gamma \alpha$ versus $\alpha$ for
the lowest instability mode, $k_x=\pi a/2 \ell$. Shown in inset is
the plot for small values of $\alpha$. \label{fig:gamma_alpha}}
\end{figure}

Denoting $c(\alpha)/\alpha+r(\alpha,k_x)$ as $\gamma(\alpha,k_x)$ we get equation (\ref{eq:G1de}) in the main text. The dependence of 
$\gamma(\alpha,k_x)$ on $\alpha$ for the main instability mode $k_x=\pi a/2 \ell$ is shown on Figure \ref{fig:gamma_alpha}.

\subsection{Calculation of $G$ up to the second order}

To calculate $G_{x,y}^2$ we need to take following integral:

\begin{equation}
 \label{eq:G1db}
 \frac{k_y^2 a}{\pi \ell} \int^{\ell/a}_0 \cos k_x\xi d\xi 
 \int^{\ell/a}_0 d\xi \ln(k_y\sqrt{(\xi-\xi')^2+\alpha^2})\cos k_x'\xi')
\end{equation}

To do this we have to split $\ln(k_y\sqrt{(\xi-\xi')^2+\alpha^2})$ into $\ln(k_y)+\frac{1}{2}\ln((\xi-\xi')^2+\alpha^2)$.
The first part of integral can be solved analytically quite easily and using that $k_x=\frac{\pi a}{2 \ell}(1+2n)$ we get following result:
\begin{equation}
 \label{eq:G1d1}
 \frac{k_y^2 a \ln{k_y}}{\pi \ell k_x^2}
\end{equation}
The second part of integral (\ref{eq:G1db}) cannot be solved analytically. 
Using numerical computation for $\alpha\ll 1$ 
we find an approximately linear dependence of this integral on $\alpha$: $a+b\alpha$. 
For $k_x=(\pi a)(2 \ell)$ this will be $-1.386+3.1\alpha$.
The full equation for $G_{x,y}$ will be:
\begin{equation}
 \label{eq:G2}
G_{x,y}^2(k_x,k_x,k_y)=\frac{a}{\ell}\left[\frac{1-\gamma(\alpha,k_x)\alpha}{\alpha} + \frac{k_y^2\ln{k_y}}{\pi k_x^2} +\frac{k_y^2}{2\pi} 
(a+b\alpha)\right]
\end{equation}
Using (\ref{eq:G2}) we can get more precise equations for (\ref{eq:de000}). The changes will be very simple
 - all terms containing $\gamma\alpha$ will transform into $(\gamma\alpha-\alpha\frac{k_y^2\ln{k_y}}{\pi 
k_x^2}-\alpha\frac{k_y^2}{2\pi}(a+b\alpha))$:
\begin{eqnarray}
  \label{eq:de03}
A_1&=&n (\gamma-\frac{k_y^2\ln{k_y}}{\pi k_x^2}-\frac{k_y^2}{2\pi}(a+b\alpha) \alpha\, ,\\ 
A_2&=&k_y^2(1+\tau A_1)+nk_x^2  +A_1 (h\tau-1)\,, \nonumber \\
A_3&=&k_y^4\tau+nk_x^2k_y^2\tau+nk_x^2(h\tau+1/n)+k_y^2(h\tau-1). \nonumber
\end{eqnarray}

\subsection{Exact expression for $h_{crit}$}

\begin{equation}
h_{crit}=\frac{2 \gamma^2 \mu_0^2 j_c^4 d^3 \kappa}{\pi^2 T^{* 2} k_x^2 C^2}\left(n + \sqrt{n^2+n} + 0.5 + \sqrt{7}\cdot 10^{-10}n \right)
\end{equation}


\begin{thebibliography}{99}
\bibitem{heatx}  Strictly speaking
$H_{\text{uni}}$ should increase at very small $E$ the same way it does in bulk
superconductors.\cite{slab,Mints81} This behavior is not reproduced in our model
since we neglected the heat flow in the $x$ direction.
This flow becomes important only in the limiting case $h_0\to 0$.
In practice, the heat removal from a thin film to the substrate usually dominates
the lateral heat diffusion in the film.
\end{thebibliography}
\end{document}